\documentclass[aps,prl,reprint,groupedaddress]{revtex4-1}

\usepackage{graphicx}

\begin{document}
\draft
\title{Dissipation-accelerated entanglement generation}
\author{Xiao-Wei Zheng$^{1}$}
\author{ Jun-Cong Zheng$^{1}$}
\author{Xue-Feng Pan$^{1}$}
\author{Li-Hua Lin$^{2}$}
\thanks{E-mail: linlh98@163.com} 
\author{Pei-Rong Han$^{2,3}$} 
\author{Peng-Bo Li$^{1}$}
\thanks{E-mail: lipengbo@mail.xjtu.edu.cn}
\address{$^{1}$School of Physics, Xi'an Jiaotong University, Xi'an 710049,\\
China\\
$^{2}$Department of Physics, Fuzhou University, Fuzhou 350108, China\\
$^{3}$School of Physics and Mechanical and Electrical Engineering, Longyan\\
University, Longyan, Fujian 364012, China}
\date{\today }

\begin{abstract}
Dissipation is usually considered a negative factor for observing quantum
effects and for harnessing them for quantum technologies. Here we propose a
scheme for speeding up the generation of quantum entanglement between two
coupled qubits by introducing a strong dissipation channel to one of these
qubits. The maximal entanglement is conditionally established by evenly
distributing a single excitation between these two qubits. When the
excitation is initially held by the dissipative qubit, the dissipation
accelerates the excitation re-distribution process for the quantum state
trajectory without quantum jumps. Our results show that the time needed to
conditionally attain the maximal entanglement is monotonously decreased as
the dissipative rate is increased. We further show that this scheme can be
generalized to accelerate the production of the W state for the three-qubit
system, where one NH qubit is symmetrically coupled to two Hermitian qubits.
\end{abstract}

\vskip0.5cm

\narrowtext

\maketitle

\section{Introduction}

\bigskip One of the most fundamental postulations of quantum mechanics is
that the state of an isolated quantum system is described by a wavefunction,
whose evolution is governed by a Hermitian Hamiltonian according to the Schr%
\"{o}dinger equation. However, any quantum system is inevitably coupled to
its surrounding environment, which acts as a meter that continuously
acquires information about the system [1]. At the zero temperature, the
environment can be regarded as a reservoir, composed of infinitely many
bosonic modes, each initially in its vacuum state [2-4]. The system's
information is gradually leaked into the reservoir by exchanging energy
quanta. The most remarkable feature of this process is that the system's
state evolution trajectory is significantly modified even if no quanta are
leaked into the environment, as a consequence of measurement backaction.

For the no-jump trajectory, the system's evolution still can be described by
the Schr\"{o}dinger equation with a non-Hermitian (NH) Hamiltonian,
featuring the competition between Hermitian and dissipative terms [5]. Such
environment-induced non-Hermiticity endows the Hamiltonian's eigenstates and
eigenenergies with striking features that are otherwise inaccessible, such
as real-to-complex spectral phase transitions and NH topology [6-9]. So far,
these NH effects have been experimentally demonstrated in distinct physical
systems [10-30]. Recently, exceptional entanglement phenomena have been
investigated in both theory and experiment in a NH spin-boson system [31].
In a very recent work [32], it was demonstrated that the entanglement
between two interacting NH qubits can be generated within a time
significantly shorter than that for two Hermitian ones. In the scheme, the
two-qubit entanglement appears as a consequence of the competition between
the intra-qubit coupling and individual driving of the qubits, and the
entanglement speedup is enabled by proximity to a higher-order exceptional
point. Each NH qubit is realized with a three-level natural or artificial
atom, where the highest and intermediate levels are used to encode the
quantum information, and the non-Hermiticity is manifested by the decay from
the intermediate to the lowest levels. This scheme is valid only when the
decaying channel from the highest to intermediate levels is neglected.
However, such a condition cannot be satisfied for general systems. For
example, the decaying rate of the levels of a normal transmon linearly
scales with the quantum number.

\bigskip We here propose an alternative scheme for exploiting
non-Hermiticity to speed up entanglement generation speedup between two
qubits. The theoretical model involves a Hermitian qubit interacting with a
decaying one by swapping coupling, which conserves the total excitation
number. The entanglement is generated in the single-excitation subspace. The
excitation, initiall possessed by the NH qubit, is distributed between the
two qubits by energy exchange. The maximal entanglement is attained when the
two qubits are equally populated. We find that the non-Hermiticity
significantly accelerate this excitation distribution process in a broad
regime, enabling the two qubits to be entangled within a time much shorter
than the dynamical timescale of the inter-qubit interaction. In distinct
contrast with the scheme of Ref. [32], our approach does not require to
encode the qubits in the highest and intermediate levels of a three-level
system; the non-Hermiticity is manifested by the decay from the higher to
lower levels of a two-level system. Futhermore, our scheme does not require
individual driving of the qubits. Due to these simplications, our scheme is
not subjected to the errors coming from the decay of the third level and
from the fluctuations of individual drives, which are inherent in the
previous scheme [33]. The idea can be directly extended to the three-qubit
system, where a NH qubit is symmetrically coupled to two Hermitian ones. The
dissipation of the NH qubit helps to speed up the generation of the
three-qubit W-type maximally entangled state in a probabilistic manner.

\section{Two-qubit entanglement speedup}

We here consider the system involving two qubits with the same frequency
interacting with each other by swapping coupling, as shown in Fig. 1a. The
non-Hermiticity of the system is manifested by the non-negligible
dissipative rate of the first qubit, denoted as $\kappa $. The decaying rate
of the second qubit is much smaller than the inter-qubit swapping rate $%
\lambda $, thereby negligible. The system dynamics is described by the
master equation%
\begin{equation}
\frac{d\stackrel{\symbol{94}}{\rho }}{dt}=-i[\stackrel{\symbol{94}}{{\cal H}}%
_{NH},\rho ]+\kappa \stackrel{\symbol{94}}{\sigma }_{1}^{-}\rho \stackrel{%
\symbol{94}}{\sigma }_{1}^{+},
\end{equation}%
where $\stackrel{\symbol{94}}{{\cal H}}_{NH}$ represents the two-qubit NH
Hamiltonian, defined as%
\begin{equation}
\stackrel{\symbol{94}}{{\cal H}}_{NH}=\lambda (\stackrel{\symbol{94}}{\sigma 
}_{1}^{+}\stackrel{\symbol{94}}{\sigma }_{2}^{-}+\stackrel{\symbol{94}}{%
\sigma }_{1}^{-}\stackrel{\symbol{94}}{\sigma }_{2}^{+})-i\frac{\kappa }{2}%
\left\vert 1\right\rangle _{1}\left\langle 1\right\vert .
\end{equation}%
Here $\stackrel{\symbol{94}}{\sigma }_{j}^{+}=\left\vert 1\right\rangle
_{j}\left\langle 0\right\vert $ and $\stackrel{\symbol{94}}{\sigma }%
_{j}^{-}=\left\vert 0\right\rangle _{j}\left\langle 1\right\vert $ with $%
\left\vert 1\right\rangle _{j}$ ($\left\vert 0\right\rangle _{j}$) denoting
the upper (lower) level of the jth qubit, and $\lambda $ is inter-qubit
coupling strength. The system evolution is a weighted mixture of infinitely
many trajectories governed by the NH Hermiltonian but interrupted by
randomly-occurring quantum jumps. For the trajectory without quantum jump,
the system state evolution is governed by $\stackrel{\symbol{94}}{{\cal H}}%
_{NH}$.

\begin{figure}
	\includegraphics[width=3.3in]{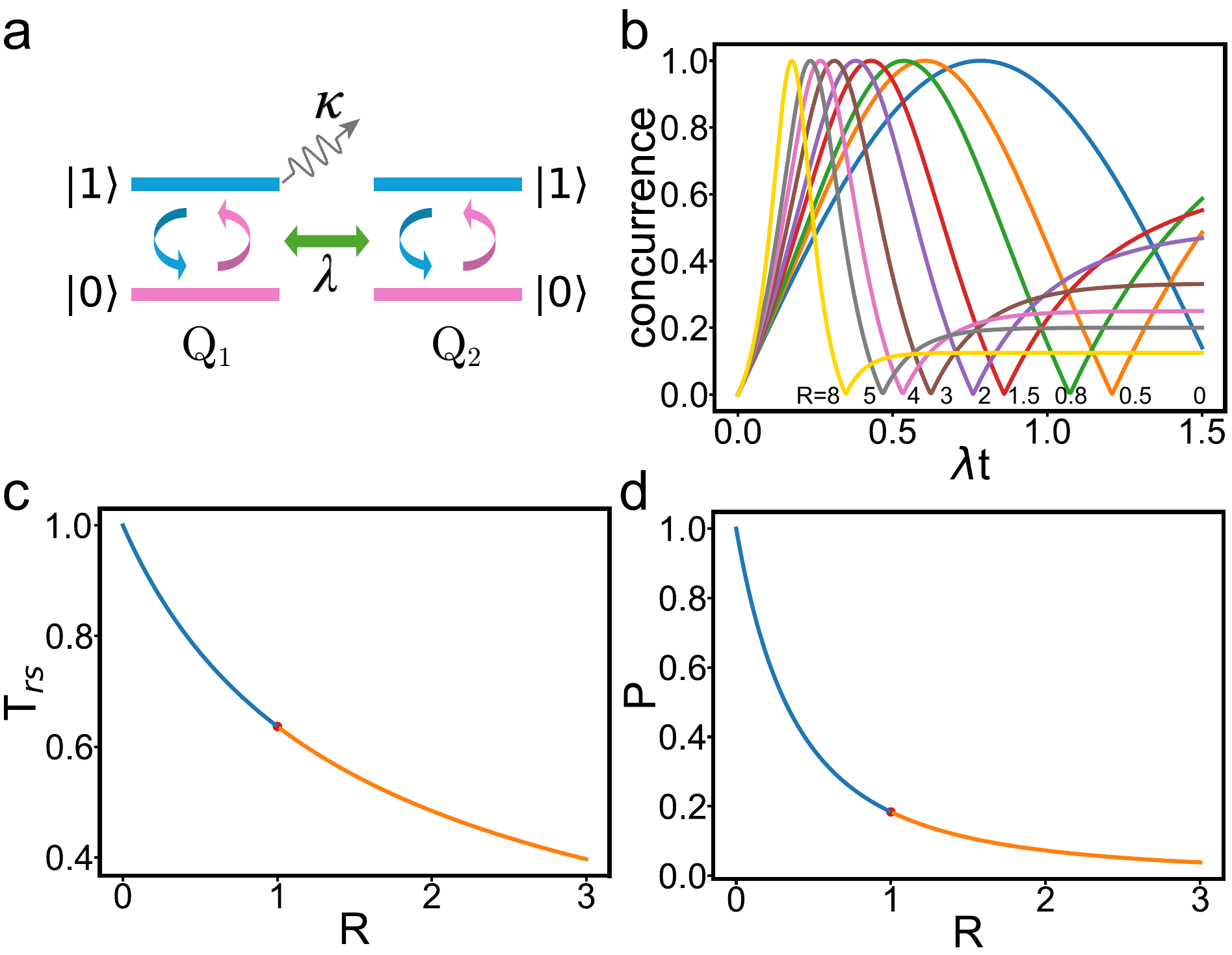} 
	\caption{(a) Sketch of the two-qubit system. The NH qubit ($Q_{1}$) and Hermitian
		qubit ($Q_{2}$) is resonantly coupled to each other with the strength $%
		\lambda $. The upper level of $Q_{1}$, $\left\vert 1\right\rangle _{1}$, has
		a non-negligible decaying rate $\kappa $. (b) Two-qubit concurrences as
		functions of $\lambda t$ for different values of $R=\kappa /4\lambda $. The
		system evolves with the initial state $\left\vert 1,0\right\rangle $ under
		the NH Hamiltonian of Eq. (1). (c) Rescaled entangling time $T_{rs}$ versus
		the rescaled dissipation rate $R$. Here $T_{rs}$ is defined as the ratio
		between the times needed to reach maximal entanglement under the NH and
		Hermitian Hamiltonians with the same inter-qubit coupling $\lambda $. The
		dot denotes the exceptional point $R=1$, where the entanglement time cannot
		be well defined. (d) Success probability for obtaining the maximally
		entangled state as a function of $R$. In the numerical simulations of (b),
		(c), and (d), the system evolves from the initial state $\left\vert
		1,0\right\rangle $ under the NH Hamiltonian of Eq. (1).}
\end{figure}

Under the NH Hamiltonian, the total excitation number of the qubits is
conserved. Consequently, the qubit Hilbert space can be devided into three
uncoupled subspaces ${\cal S}_{0}=\left\{ \left\vert
0_{1},0_{2}\right\rangle \right\} $, ${\cal S}_{1}=\left\{ \left\vert
1,0\right\rangle ,\left\vert 0,1\right\rangle \right\} $, and ${\cal S}%
_{2}=\left\{ \left\vert 1,1\right\rangle \right\} $. In ${\cal S}_{1}$, the
system has a pair of eigenenergies, given by 
\begin{equation}
E_{\pm }=-i\kappa /4\pm \Omega .
\end{equation}%
where $\Omega =\sqrt{\lambda ^{2}-\kappa ^{2}/16}$. The corresponding
eigenstates are 
\begin{equation}
\left\vert \Phi _{\pm }\right\rangle ={\cal N}_{\pm }(E_{\pm }\left\vert
1_{1},0_{2}\right\rangle +\lambda \left\vert 0_{1},1_{2}\right\rangle ),
\end{equation}%
where ${\cal N}_{\pm }=(\left\vert E_{\pm }\right\vert ^{2}+\lambda
^{2})^{-1/2}$. Suppose that the system is initially in the state $\left\vert
\psi (0)\right\rangle =\left\vert 1_{1},0_{2}\right\rangle $. Then the
quantum state evolution associated with the no-jump trajectory can be
rewritten as%
\begin{equation}
\left\vert \psi (t)\right\rangle ={\cal N}({\cal A}\left\vert
1,0\right\rangle -i{\cal B}\left\vert 0,1\right\rangle ),
\end{equation}%
where ${\cal N}=({\cal A}^{2}+{\cal B}^{2})^{-1/2}$. When $R=\kappa
/(4\lambda )<1$, the coefficients ${\cal A}$ and ${\cal B}$ are given by 
\begin{eqnarray*}
{\cal A} &=&\cos (\Omega t)-\frac{\kappa }{4\Omega }\sin (\Omega t), \\
{\cal B} &=&\frac{\lambda }{\Omega }\sin (\Omega t).
\end{eqnarray*}%
For $R>1$, the trigonometric functions are replaced by the corresponding
hyperbolic functions, with $\Omega =\sqrt{\kappa ^{2}/16-\lambda ^{2}}$. The
two-qubit entanglement is quantified by the concurrence, given by [33]%
\begin{equation}
{\cal E}=2{\cal N}^{2}{\cal AB}\text{.}
\end{equation}

To quantify to what extent the entanglement generation is speeded up by the
non-Hermiticity, we define the rescaled entangling time, $T_{rs}=T/\tau $,
where $T$ is the shortest time needed to conditionally produce the two-qubit
maximally entangled state for the NH system, and $\tau =\pi /4\lambda $ is
the corresponding time for the Hermitian system. For $R<1$, the rescaled
entangling time is given by%
\begin{equation}
T_{rs}=\frac{4}{\pi }\frac{1}{\sqrt{1-R^{2}}}\arctan \sqrt{\frac{1-R}{1+R}}.
\end{equation}%
When $R>1$, $T_{rs}$ is%
\begin{equation}
T_{rs}=\frac{4}{\pi }\frac{1}{\sqrt{R^{2}-1}}arc\tanh \sqrt{\frac{R-1}{R+1}},
\end{equation}

To quantitatively confirm the non-Hermiticity-enabled entanglement speedup,
in Fig. 1b we present the two-qubit concurrence, associated with the no-jump
trajectory, as a function of $\lambda t$ and $R$. The result clearly shows
that the concurrence ${\cal E}$ exhibits a non-monotonous behavior, and can
reach the maximum 1 no matter when $R<1$ or $R>1$. The time needed to reach
the maximal entanglement (${\cal E}=1$) is monotonously decreased with the
increase of $\kappa $. When $R>1$, after reaching the maximum ${\cal E}$
drops to zero at the time $t=G^{-1}arc\tanh (4G/\kappa )$, and then
monotonously tends towards ${\cal E}_{t\rightarrow \infty }=2x/(1+x^{2})$,
where $x=(G-\kappa /4)/\lambda $. For $R\gg 1$, ${\cal E}_{t\rightarrow
\infty }\simeq 1/(2R)$. Fig. 1c display the rescaled entangling time $T_{sc}$
versus the ratio $R$. The results clearly show that the entanglement
generation is indeed speeded up by the non-Hermiticity, and $T_{sc}^{2q}$
has a maximum $1$ when $\kappa =0$. The result can be interpreted as
follows. The maximal entanglement is reached when the populations of $%
\left\vert 1,0\right\rangle $ and $\left\vert 0,1\right\rangle $, $P_{1,0}$
and $P_{0,1}$, are balanced. For the unitary evolution, this is achieved by
coherent excitation transfer. When the excitation is initially populated in
the NH qubit, the leakage of its population from the NH qubit to the
environment help to increase the ratio $P_{0,1}/P_{1,0}$ from 0 to 1.

This entanglement dynamics is in stark contrast with that presented in Re.
[32], where the system starts with the initial state $\left\vert
0,1\right\rangle $. For such an initial state, the system evolution
associated with the no-jump trajectory is given by [32]%
\begin{equation}
\left\vert \psi ^{\prime }(t)\right\rangle ={\cal N}^{\prime }({\cal A}%
^{\prime }\left\vert 0,1\right\rangle -i{\cal B}\left\vert 1,0\right\rangle
),
\end{equation}%
where ${\cal A}^{\prime }=\cos (\Omega t)+\frac{\kappa }{4\Omega }\sin
(\Omega t)$ and $\cosh (Gt)+\frac{\kappa }{4G}\sinh (Gt)$ for $R<1$ and $R>1$%
, respectively. The two-qubit entanglement is ${\cal E}^{\prime }=2{\cal N}%
^{\prime 2}{\cal A}^{\prime }{\cal B}$. For $R<1$, the rescaled entangling
time is%
\begin{equation}
T_{rs}^{\prime }=\frac{4}{\pi }\frac{1}{\sqrt{1-R^{2}}}\arctan \sqrt{\frac{%
1+R}{1-R}}.
\end{equation}%
Contrary to the case with the initial state $\left\vert 1,0\right\rangle $, $%
T_{rs}^{\prime }$ increases with $R$. This is due to the fact that the
dissipation plays a negative role in the transfer of the excitation from the
Hermitian qubit to the NH qubit, so that it takes a longer time to balance $%
P_{1,0}$ and $P_{0,1}$. When $R>1$, ${\cal E}$ is monotonously increased,
tending to a fixed value, which is smaller than 1 and is decreased when R
increases. The results imply that the dissipation can speed up entanglement
only when excitation is initially held by the NH qubit.

It is worthwhile to investigate to what extent the success probability of
producing the maximally entangled state is reduced with the increase of the
dissipation rate. When $R<1$, this probability is 
\begin{equation}
{\cal P}=2\frac{1}{1-R^{2}}e^{-\kappa T/2}\sin ^{2}\left( \Omega T\right) .
\end{equation}%
For $R>1$, ${\cal P}$ is 
\begin{equation}
{\cal P}=2\frac{1}{R^{2}-1}e^{-\kappa T/2}\sinh ^{2}\left( \Omega T\right) .
\end{equation}%
Fig. 1d presents this probability as a function of $R$. The result implies
that, for a given inter-qubit coupling strength, the time needed to achieve
the maximal entanglement is shortened at the price of the decrease of the
success probability.

\section{Three-qubit entanglement speedup}

\begin{figure}
	\includegraphics[width=3.3in]{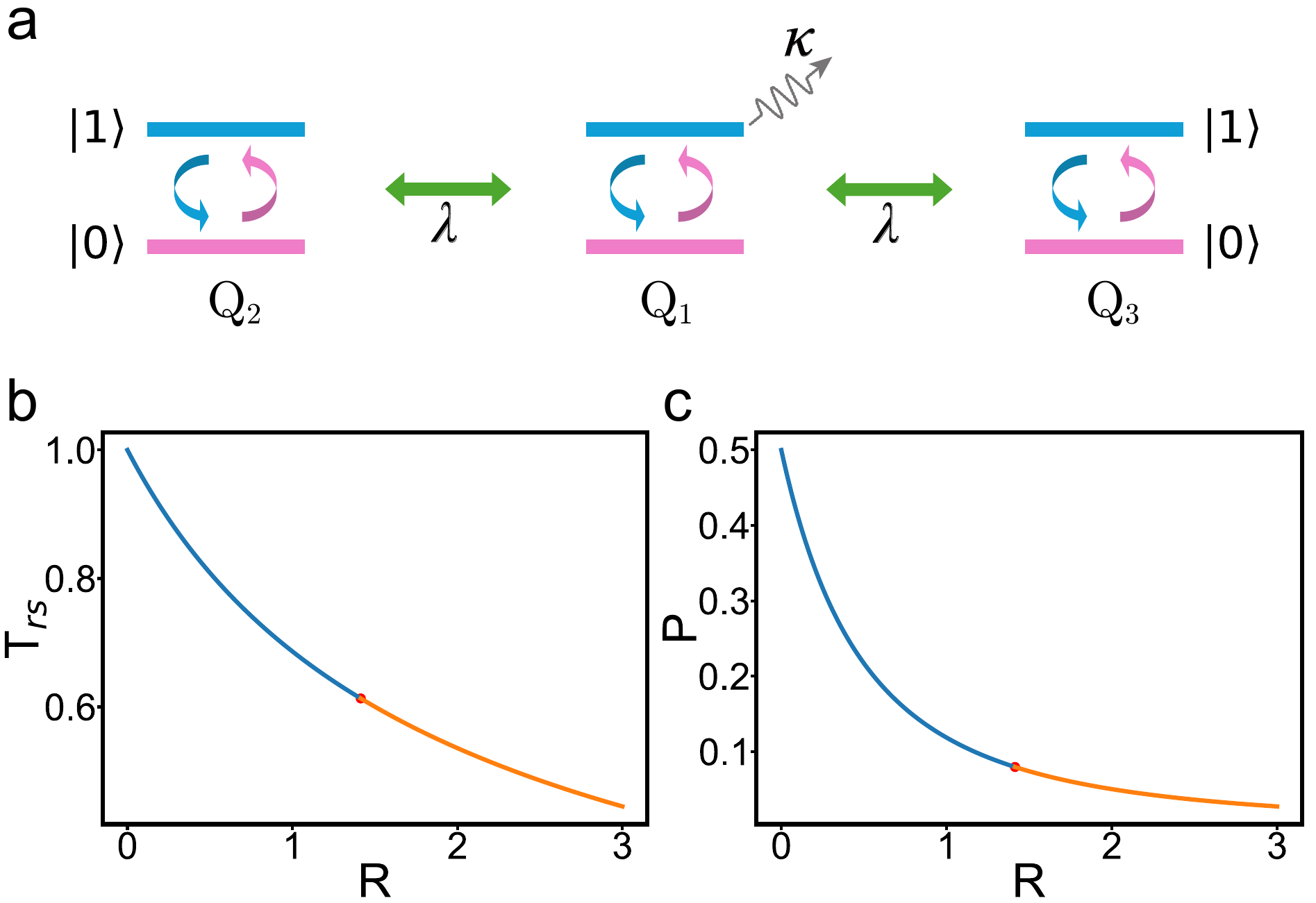} 
	\caption{(a) Sketch of the three-qubit system. One NH qubit ($Q_{1}$) is
		symmetrically coupled to two Hermitian qubits, $Q_{2}$ and $Q_{3}$, with the
		strength $\lambda $. (b) Rescaled entangling time $T_{rs}$ versus the $R$.
		Here $T_{rs}$ is defined as the ratio between the times needed to reach the
		W-type maximally entangled state under the NH and Hermitian Hamiltonians
		with the same inter-qubit coupling $\lambda $. The system is initially in
		the state $\left\vert 1,0,0\right\rangle $ and evolves under the three-qubit
		NH Hamiltonian. (c) Success probability for obtaining the W-type maximally
		entangled state as a function of $R$.}
\end{figure}

We note that the scheme can be directly generalized to realize
dissipation-based three-qubit entanglement accelaration. The system involves
two Hermitian qubits, each of which is coupled to a NH qubit with the
coupling strength $\Omega $, as shown in Fig. 2a. For the no-jump evolution
trajectory, the system dynamics is governed by the NH Hamiltonian%
\begin{equation}
\stackrel{\symbol{94}}{{\cal H}}_{NH}=\lambda \lbrack \stackrel{\symbol{94}}{%
\sigma }_{1}^{+}(\stackrel{\symbol{94}}{\sigma }_{2}^{-}+\stackrel{\symbol{94%
}}{\sigma }_{3}^{-})+H.c.]-i\frac{\kappa }{2}\left\vert 1\right\rangle
_{1}\left\langle 1\right\vert .
\end{equation}%
In the single-excitation subspace $\left\{ \left\vert 1,0,0\right\rangle
,\left\vert 0,1,0\right\rangle ,\left\vert 0,0,1\right\rangle \right\} $,
the system has three eigenenergies, given by 
\begin{eqnarray}
E_{0} &=&0, \\
E_{\pm } &=&-i\kappa /4\pm \Lambda ,  \nonumber
\end{eqnarray}%
where $\Lambda =\sqrt{2\lambda ^{2}-\kappa ^{2}/16}$. The corresponding
eigenstates are

\begin{eqnarray}
\left\vert \Phi _{0}\right\rangle  &=&\frac{1}{\sqrt{2}}(\left\vert
0,1,0\right\rangle -\left\vert 0,0,1\right\rangle ), \\
\left\vert \Phi _{j}\right\rangle  &=&{\cal N}_{\pm }(E_{\pm }\left\vert
1,0,0\right\rangle +\sqrt{2}\lambda \left\vert \phi _{b}\right\rangle , 
\nonumber
\end{eqnarray}%
where $\left\vert \phi _{b}\right\rangle =\frac{1}{\sqrt{2}}(\left\vert
0,1,0\right\rangle +\left\vert 0,0,1\right\rangle )$. When the system is
initially in the state $\left\vert \psi (0)\right\rangle =\left\vert
1,0,0\right\rangle $, the state evolution associated to the no-jump
trajectory is%
\begin{equation}
\left\vert \Psi (t)\right\rangle ={\cal M}({\cal C}\left\vert
1,0,0\right\rangle -i{\cal D}\left\vert \phi _{b}\right\rangle ),
\end{equation}%
where ${\cal M}=\sqrt{{\cal C}^{2}+{\cal D}^{2}}$. When $R<\sqrt{2}$, the
coefficients ${\cal C}$ and ${\cal D}$ are given by 
\begin{eqnarray}
{\cal C} &=&\cos (\Lambda t)-\frac{\kappa }{4\Lambda }\sin (\Lambda t), \\
{\cal D} &=&\frac{\lambda }{\Lambda }\sin (\Lambda t).  \nonumber
\end{eqnarray}%
For $R>\sqrt{2}$, the trigonometric functions are replaced by the
corresponding hyperbolic functions, with $\Lambda =\sqrt{\kappa
^{2}/16-2\lambda ^{2}}$.

When $\left\vert {\cal C}\right\vert =\left\vert {\cal D}\right\vert /\sqrt{2%
}$, $\left\vert \Psi (t)\right\rangle $ corresponds to the three-qubit
W-type maximally entangled state [34]%
\begin{equation}
\left\vert \Psi _{w}\right\rangle =(e^{i\phi }\left\vert 1,0,0\right\rangle
\ +\left\vert 0,1,0\right\rangle +\left\vert 0,0,1\right\rangle )/\sqrt{3}.
\end{equation}%
For such a W state, any two qubits have a concurrence of $2/3$. When $R<%
\sqrt{2}$, the rescaled three-qubit W-state entangling time is given by%
\begin{equation}
T_{rs}=\frac{\sqrt{2}}{\sqrt{2-R^{2}}\arccos (1/\sqrt{3})}\arctan \frac{%
\sqrt{2-R^{2}}}{1+R}.
\end{equation}%
Here $T_{rs}$ is defined as the ratio between the W-state generation times
for the NH and Hermitian systems, $T_{rs}=T/\tau $, where $\tau =\frac{1}{%
\sqrt{2}\lambda }\arccos (1/\sqrt{3})$. For $R>\sqrt{2}$, the rescaled
W-state entangling time is%
\begin{equation}
T_{rs}=\frac{\sqrt{2}}{\sqrt{R^{2}-2}arc\cosh (1/\sqrt{3})}arc\tanh \frac{%
\sqrt{R^{2}-1}}{1+R}.
\end{equation}%
The rescaled W-state generation time $T_{sc}$, as a function of $R$, is
presented in Fig. 2b, which confirms that $T_{sc}$ is smaller than 1 when $%
\kappa \neq 0$, verifying the non-Hermiticity-enabled acceleration of the W
state generation. The result demonstrates that the time needed to reach the
W-type maximally entangled state monotonously decreases with the increase of 
$\kappa $ for $\lambda <\kappa /4\sqrt{2}$. When $R<\sqrt{2}$, the success
propability for producing the W-type maximally entangled state is 
\begin{equation}
{\cal P}=\frac{3}{2(2-R^{2})}e^{-\kappa T/2}\sin ^{2}(\Lambda T).
\end{equation}%
For $R>\sqrt{2}$, ${\cal P}$ is given by 
\begin{equation}
{\cal P}=\frac{3}{2(R^{2}-2)}e^{-\kappa T/2}\sinh ^{2}(\Lambda T).
\end{equation}%
which decreases with the increase of $R$, as shown in Fig. 2c.

\section{Conclusion}

In conclusion, we have presented a scheme for exploiting dissipation to
speed up entanglement evolution for one NH qubit and one Hermitian qubit,
interacting with each other via swapping a single excitation. When the
excitation is initially possessed by the NH qubit, the system can
conditionally evolve to a maximally entangled state within a time shorter
than that needed for the Hermitian system with the same inter-qubit coupling
strength. Neither does the three-level configuration nor individual driving
of the qubits are required. The time needed to achieve the maximal
entanglement monotonously decreases with the increase of the dissipation
rate at the price of decreasing success probability. We further show that
scheme can be extended to the three-qubit case, where one NH qubit initially
in its excited state is coupled to two Hermitian qubits initially in their
ground states. For the no-jump evolution trajectory, the time required to
produce the W state is reduced by the dissipation of the NH qubit.

This work was supported by the National Natural Science Foundation of China
under Grant No. 12474356.

\end{document}